\documentclass[11pt,twoside]{article}
\usepackage{asp2004}
\usepackage{epsf}
\usepackage{epsfig}
\usepackage{lscape}

\begin{document}
\title{Polarimetric standardization} 
\author{E.\ Landi Degl'Innocenti$^1$, 
        S.\ Bagnulo$^2$, \&
        L.\ Fossati$^{2,3}$}
\affil{$^1$\ Dipartimento di Astronomia e Scienza dello Spazio,
           Universit\`{a} degli Studi di Firenze,
           Largo Enrico Fermi 2, I-50125 Firenze, Italy \\
       $^2$\ European Southern Observatory,
           Alonso de Cordova 3107, Vitacura, Santiago, Chile \\
       $^3$\ Institut fuer Astronomie, Universitaet Wien,
           Tuerkenschanzstrasse 17, A-1180 Wien, Austria
       }

\begin{abstract}
The use of polarimetric techniques is nowadays widespread among solar
and stellar astronomers. However, notwithstanding the recommandations
that have often been made about the publication of polarimetric
results in the astronomical literature, we are still far from having a
standard protocol on which to conform. In this paper we review the
basic definitions and the physical significance of the Stokes
parameters, and we propose a standardization of the measurement of
polarized radiation.
\end{abstract}

\section{Introduction}
The polarization properties of a radiation beam can be described in
several different ways but have ultimately to rely on the
specification of four independent quantities. Particularly suitable
from this point of view are the so-called Stokes parameters which were
introduced in the scientific literature by George Gabriel Stokes
\cite{Stokes52} and which are commonly referred to, in modern
notations, with the symbols $I$, $Q$, $U$, and $V$. It is possible to
give two different definitions for the Stokes parameters of a beam of
electromagnetic radiation propagating along a given
direction. According to the first one, the Stokes parameters are expressed
as statistical averages of bilinear products of the components of the
electric field along two perpendicular axes, both perpendicular to the
direction of propagation of the radiation (see
Sect.~3).  This basic definition is also used in
practice by radio-astronomers for expressing their measurements of the
Stokes parameters.  The second definition, mostly used by optical
astronomers, is an operational one that makes use of the concept of
ideal filters (see Sect.~7). Obviousy, the two
definitions have to be consistent.

\section{The choice of the reference system}
Before introducing these two definitions, it is fundamental to point
out that both of them need the preliminary choice of a {\it reference
direction} pertaining to the plane perpendicular to the direction of
propagation of the radiation. To introduce the first definition of the
Stokes parameters (the one based on bilinear products of the
components of the electric field), we refer to a {\it right-handed}
coordinate system, ($x,y,z$), with the $z$-axis directed along the
direction of propagation and the $x$-axis directed along the reference
direction. In astronomical observations, the reference direction can
be thought of as the great circle pertaining to the celestial sphere
and passing through the observed object. Though the choice of the
great circle is, in principle, arbitrary, it has become customary,
at least in night-time astronomy, to choose the reference direction as
the celestial meridian passing through the observed object. In solar
observations this convention is not always followed since there are,
in general, more physical directions to choose, like the tangent to
the solar limb for prominence observations, or the direction passing
through the observed point and the center of the solar disk for
sunspots observations. In polarimetric measurements of other solar
system objects (e.g., asteroids) it is common to adopt as reference
direction the great circle passing through the object itself and the
Sun, or, better choice, the perpendicular to it (see Sect.~6). 

\section{The basic definition}\label{Sect_Bilinear}
According to Maxwell
equations, the electric field vector of the radiation beam lies in the
$(x,y)$ plane so that it is described by only the two components $E_x$
and $E_y$. In a fixed point of space along the direction of
propagation, at the entrance of the telescope for instance, these two
components of the electric field are described by the functions of
time, $E_x(t)$ and $E_y(t)$, for which we can introduce the Fourier
transforms, ${\cal E}_x(\omega)$ and ${\cal E}_y(\omega)$, according
to the usual definitions

\begin{equation}
\begin{array}{rcl}
{\cal E}_x(\omega) &=& \int_{-\infty}^{\infty} E_x(t) \, {\rm e}^{-{\rm i} \omega t}
\, {\rm d} t \\
{\cal E}_y(\omega) &=& \int_{-\infty}^{\infty} E_y(t) \, {\rm e}^{-{\rm i} \omega t}
\, {\rm d} t \; \;. \\
\end{array}
\label{Eq_Fourier}
\end{equation}
The Stokes parameters of the radiation beam at the angular frequency $\omega$
are defined by the expressions
\begin{equation}
\begin{array}{rcl}
   I(\omega) &=& \phantom{{\rm i}}
                 k \, [ \langle {\cal E}_x(\omega)^* {\cal E}_x(\omega) \rangle + 
   \langle {\cal E}_y(\omega)^* {\cal E}_y(\omega) \rangle ]  \\ [2mm]
   Q(\omega) &=& \phantom{{\rm i}}
                 k \, [ \langle {\cal E}_x(\omega)^* {\cal E}_x(\omega) \rangle - 
   \langle {\cal E}_y(\omega)^* {\cal E}_y(\omega) \rangle ] \\ [2mm]
   U(\omega) &=& \phantom{{\rm i}}
                 k \, [ \langle {\cal E}_x(\omega)^* {\cal E}_y(\omega) \rangle + 
   \langle {\cal E}_y(\omega)^* {\cal E}_x(\omega) \rangle ] \\ [2mm]
   V(\omega) &=& {\rm i} k \, [ \langle {\cal E}_x(\omega)^* {\cal E}_y(\omega) 
   \rangle - \langle {\cal E}_y(\omega)^* {\cal E}_x(\omega) \rangle ] \\
\end{array}
\label{Eq_Basic}
\end{equation}
where $k$ is a positive constant, the symbol $*$ means complex conjugate, and
the symbol $\langle \cdots \rangle$ means a statistical average which
is implicit in the process of measurement.

There are some risks of ambiguity in the definition of Stokes
parameters. One of course is the choice of the handedness of the
reference system. We have adopted a right handed reference system, but
adopting a left handed system will change the signs of $U$ and $V$.

There is a much more subtle ambiguity arising from the definition of
Fourier transform. In our definitions of Eq.~(\ref{Eq_Basic}) we have
implicitely adopted for the Fourier transform the representation of
Eq.~(\ref{Eq_Fourier}).  Introducing the alternative definition with
the substitution
\[
 {\rm e}^{-{\rm i} \omega t} \rightarrow {\rm e}^{{\rm i} \omega t} 
   \;\; 
\]
namely, defining
\begin{equation}
\begin{array}{rcl}
   {\cal E}'_x(\omega) &=& \int_{-\infty}^{\infty} E_x(t) \, {\rm e}^{{\rm i} \omega t} \, {\rm d} t \\[2mm]
   {\cal E}'_y(\omega) &=& \int_{-\infty}^{\infty} E_y(t) \, {\rm e}^{{\rm i} \omega t} \, {\rm d} t \\
\end{array}
\end{equation}
one obviously has, being $E_x(t)$ and $E_y(t)$ real,
\[
\begin{array}{rcl}
   {\cal E}'_x(\omega) &=& {\cal E}_x(\omega)^* \\
   {\cal E}'_y(\omega) &=& {\cal E}_y(\omega)^* \\ 
\end{array}
\]
so that the sign of the $V$ Stokes parameter changes when the definition of 
the Fourier transform is changed from Eq.~(1) to Eq.~(3). In practice,
this means that when adopting Eq.~(2) for the definition of the Stokes
parameters, one has to be aware that such definition implies a convention
about the choice of positive or negative time in the definition of the
Fourier transform. 

Finally, we want to comment on the use of an arbitrary constant in
front of the bilinear expressions that defines the Stokes parameters.
As a matter of fact, we always use relative quantities, that is, we
normalise $Q$, $U$, and $V$ to the intensity, and we consider the reduced
Stokes parameters $P_Q$, $P_U$, and $P_V$ defined by
\begin{equation}
\begin{array}{rcl}
P_Q &=& Q/I \\
P_U &=& U/I \\
P_V &=& V/I \\
\end{array}
\label{Eq_PQ_PU}
\end{equation}
so the value of the constant $k$ can be left undefined.

\section{The physical significance of the Stokes parameters}\label{Sect_Phys}
According to the definition given by Eqs.~(\ref{Eq_Basic}), Stokes $Q$
is the difference between the amount of photons whose electric field
oscillates along the reference direction and the amount of photons
whose electric field oscillates in the direction perpendicular to it;
Stokes $U$ is the difference between the amount of photons whose
electric field oscillates at $45^\circ$ and the amount of photons
whose electric field oscillates at $135^\circ$ with respect to the
reference direction (the angles are reckoned counterclockwise from the
reference direction looking at the source -- see Fig.~1).

\begin{figure}
\begin{center}
\plotfiddle{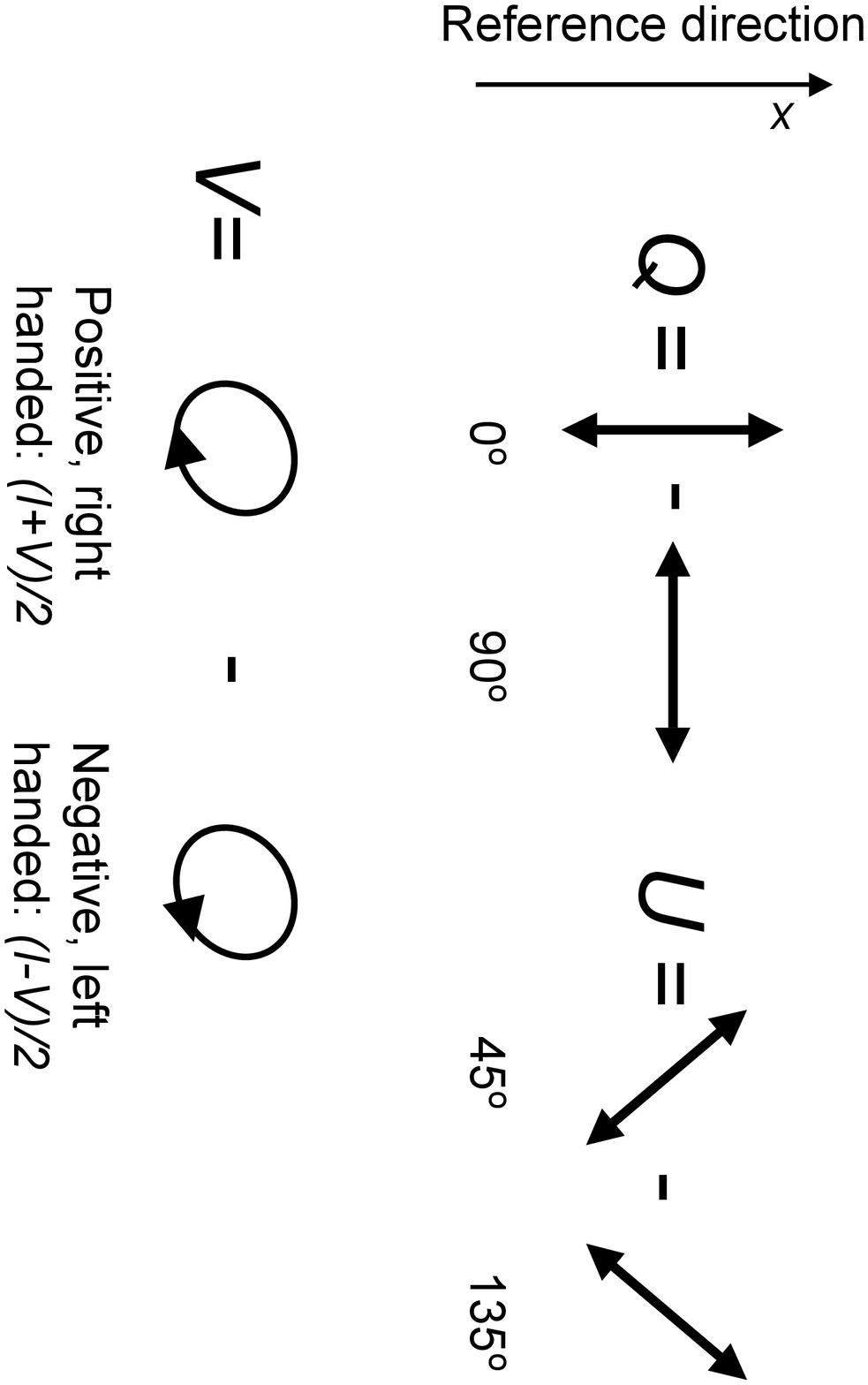}{5cm}{90}{40}{40}{180}{-60}
\end{center}
\caption{Geometrical visualization of the definition of Stokes parameters.
The observer is looking toward the source.}
\label{Fig_Def}
\end{figure}

Stokes $V$ is given by the so called right handed circular polarization
minus the left handed circular polarization, that are so defined: in a
fixed point of space, the tip of the electric field vector carried by
a beam having positive circular polarization rotates clockwise, as
seen by an observer looking at the source of radiation. Viceversa, the
tip of the electric field vector carried by a beam having negative
circular polarization rotates counterclockwise, as seen by an observer
looking at the source (see Fig.~\ref{Fig_Def}).

It is possible to show that Stokes parameters give information about
the ellipse drawn by the tip of the electric vector in the plane
perpendicular to the direction of wave propagation.  For the sake of
simplicity, one can consider the case of a monochromatic wave, which
is completely polarized. More generally, we remind
that any radiation beam can be considered as the incoherent
superposition of an unpolarized beam and a beam which is totally
polarized (Chandrasekhar theorem). The tip of the electric field
associated with this beam draws, in a fixed point of space, an
ellipse. The major and minor semi-axes of this ellipse, $a$ and $b$,
(expressed in non dimensional units), and the tilt angle, $\theta$,
that the major semi-axis forms with the
reference direction $x$ (reckoned counterclockwise -for an observer
looking at the source- from the reference direction) are connected
to the reduced Stokes parameters of the polarized component by the
equations
\begin{equation}
\begin{array}{rcl}
          a &=& \frac{1}{2}\left[(1-P_V)^{1/2} + (1+P_V)^{1/2}\right] \\ [2mm]
          b &=& \frac{1}{2}\vert(1-P_V)^{1/2} - (1+P_V)^{1/2}\vert    \\ [2mm]
    P_U/P_Q &=& \tan\left(2\theta\right)                              \\
\end{array}
\label{Eq_Geometry}
\end{equation}
where $\vert z \vert$ means absolute value of $z$.
Stokes $V$ is connected to the eccentricity of the
polarization ellipse, and Stokes $Q$ and $U$ tell us how the
polarization ellipse is oriented. If linear polarization is zero, the
polarization ellipse degenerates in a circle. If circular polarization
is zero, the polarization ellipse degenerates in a segment.

\section{Linear polarization and position angle}
Linear polarization is often expressed also in terms of $P_{\rm L}$ and
$\theta$, where $P_{\rm L}$ is the fraction of linearly polarized radiation,
and $\theta$ is the angle of maximum polarization, i.e., the angle
that the major axis of the polarization ellipse forms with the
$x$-axis of the reference system reckoned with the usual convention:
\begin{equation}
\begin{array}{rcl}
P_{\rm L} &=& \sqrt{P_Q^2 +P_U^2}\\
P_Q &=& P_{\rm L}\,\cos(2\theta) \\
P_U &=& P_{\rm L}\,\sin(2\theta) \\
\end{array}
\label{Eq_PT}
\end{equation}
The use of $P_{\rm L}$ and $\theta$ is widespread in astronomy because
it allows the graphical superposition of linear polarization results
to the image of the observed object. Note that from
Eq.~(\ref{Eq_PT}) one may be tempted to calculate the position
angle $\theta$ as $(1/2)\,\arctan(P_U/P_Q)$.  This is incorrect, as can
be seen with the following example: consider the case of $\theta = 67.5^\circ$ i.e.,
a polarization ellipse with its major axis tilted at an angle of
$67.5^\circ$ with respect to the reference $x$-axis. From
Eq.~(\ref{Eq_PT}), we obtain $P_Q = -P_{\rm L} / \sqrt{2}$ and
$P_U = P_{\rm L} / \sqrt{2}$.  By blindly computating
$(1/2)\,\arctan(P_U/P_Q)$ one will find an ellipse tilted at $-22.5^\circ$, i.e.,
perpendicular to the original, correct one. The reason for that
inconsistency is that arctan is a function defined between $-90^\circ$
and $90^\circ$. The correct expression for obtaining
the position angle (apart from inessential multiples of $\pi$) is the
following one:
\begin{equation}
\begin{array}{rcll}
\theta &=& +\frac{1}{4} \pi & \mathrm{if} \ P_Q = 0 \ \mathrm{and}\ \ P_U > 0 \\
\theta &=& -\frac{1}{4} \pi & \mathrm{if} \ P_Q = 0 \ \mathrm{and}\ \ P_U < 0 \\
\theta &=&  \frac{1}{2} \arctan \left(\frac{P_U}{P_Q} \right) 
                            & \mathrm{if} \ P_Q > 0                         \\
\theta &=&  \frac{1}{2} \arctan \left(\frac{P_U}{P_Q} \right) + \frac{\pi}{2} 
                            & \mathrm{if} \ P_Q < 0 \\
\end{array}
\label{Eq_Theta}
\end{equation}
Alternatively, one can substitute Eqs.~(\ref{Eq_Theta}) by the unique formula, 
suggested by V.\ Bommier (private communication)
\begin{equation}
\theta = {1 \over 2} \, {\rm sign}(P_U) \,
\arccos \left(\frac{P_Q}{\sqrt{P_Q^2 + P_U^2}} \right) \;\; . 
\label{Eq_Theta_Bomm}
\end{equation}

\section{Transforming Stokes parameters into a new reference system}\label{Sect_Trans}
A further point concerns the transformation law for the Stokes
parameters when changing the reference direction. If the new reference
direction is obtained from the old one by a counterclockwise (looking
at the source) rotation of an angle $\chi$, by means of
Eq.~(\ref{Eq_Basic}) one can easily show that the new reduced Stokes
parameters, $(P_Q',P_U',P_V')$, are connected to the old ones,
$(P_Q,P_U,P_V)$, by the equations
\begin{equation}
\begin{array}{rcl}
P_Q' &=& \cos \, (2 \chi) \, P_Q + \sin \, (2 \chi) \, P_U \\
P_U' &=& - \sin \, (2 \chi) \, P_Q + \cos \, (2 \chi) \, P_U \\
P_V' &=& P_V \\
\end{array}
\label{Eq_Trans}
\end{equation}
Equations~(\ref{Eq_Trans}) show that the linear polarization Stokes
parameters, $Q$ and $U$, transform between each other when the
reference direction is changed, whence the importance, as already
remarked, of clearly specifying the chosen reference direction when
publishing a result concerning linear polarization. The same equations
also show the invariance of the quantity $P_{\rm L}$ under a rotation
of the reference direction, and the obvious fact that a rotation of
the reference direction of $180^\circ$ leaves all the Stokes
parameters invariant. This last fact means that when choosing the
reference direction it is not necessary to specify ``an arrow'' on the
direction itself. If, for instance, one chooses as reference direction
the celestial meridian passing through the observed object, it does
not matter whether the $x$-axis of the system $(x,y,z)$, on which the
electric field components are defined, is directed towards the North
or the South pole.

It should be noted that Eq.\,(\ref{Eq_Trans}) says that the
position angle $\theta$ repres\-ents the angle by which one should
rotate the reference direction (in the counterclockwise direction,
looking at the source), in order to end up with a positive $P_Q$ value
(which indeed turns up to be equal to $P_{\rm L}$) and a null $P_U$
value.

As an example of the use of Eq.~(\ref{Eq_Trans}), let us consider the
polarization of a solar system object measured in a system that has
the reference direction ($x$-axis) aligned to the celestial meridian
passing through the observed object. We want to transform Stokes
parameters into a system that has the reference direction
perpendicular to the great circle passing through the object and the
Sun. We have to calculate $P_Q'$, $P_U'$, $P_V'$ via
Eqs.~(\ref{Eq_Trans}), setting $\chi = \Phi + \pi/2$, where $\Phi$ is
the angle between between the celestial meridian passing through the
object and the great circle passing through the Sun and the object.
The angle $\Phi$ is related to the coordinates of the object and of
the Sun through the relationship
\begin{equation}
\sin \delta_{\rm T} \cos(\alpha_\odot - \alpha_{\rm T}) = 
\tan \delta_\odot   \cos\delta_{\rm T} - \sin(\alpha_\odot -
\alpha_{\rm T}) \frac{1}{\tan\Phi}
\label{Eq_Solar_System}
\end{equation}
where $(\alpha_\odot,\delta_\odot)$, and $(\alpha_{\rm T},\delta_{\rm T})$ are the
right ascension and declination of the Sun, and of the observed object, respectively.
Alternatively, one can directly calculate
\begin{equation}
\begin{array}{rcl}
P_Q' &=&             \frac{{\cal B}^2- {\cal C}^2}{1 - {\cal A}^2}\, P_Q
         -  \frac{2\,{\cal B}{\cal C}}{1 - {\cal A}^2}\, P_U \\ [2mm]
P_U' &=&    \frac{2\,{\cal B}{\cal C}}{1 - {\cal A}^2}\, P_Q +
             \frac{{\cal B}^2- {\cal C}^2}{1 - {\cal A}^2}\, P_U \\
\end{array}
\end{equation}
where
\begin{equation}
\begin{array}{rcl}
{\cal A} &=& \sin \delta_\odot \sin \delta_{\rm T} + 
             \cos \delta_\odot \cos \delta_{\rm T} \cos(\alpha_\odot - \alpha_T) \\[2mm]
{\cal B} &=& \cos \delta_\odot \sin (\alpha_\odot - \alpha_{\rm T}) \\[2mm]
{\cal C} &=& \sin \delta_\odot \cos \delta_{\rm T} -
             \cos \delta_\odot \sin \delta_{\rm T} \cos(\alpha_\odot - \alpha_{\rm T})\;. \\
\end{array}
\end{equation}
This transformation is meaningfull in polarimetry of solar system
objects because in the new reference system, Stokes~$Q$ represents
the flux perpendicular to the plane Sun-Object-Earth (commonly
referred to as the \textit{scattering plane}) minus the flux parallel to
that plane.

\section{Operational definition}\label{Sect_Operational}
In optical polarimetry, it has become customary to give an alternative
definition of the Stokes parameters in terms of ideal filters for
linear and circular polarization (Shurcliff 1962). This requires the
preliminary definitions of ideal filters for linear and circular
polarization.

\subsection{An ideal filter for linear polarization}
An ideal filter for linear polarization (called a {\it polarizer}) is
a device that can be interposed along a radiation beam and which, by
definition, is totally transparent to the component of the electric
field along a given direction, perpendicular to the direction of
propagation (the transmission axis of the polarizer), and totally
opaque to the component of the electric field in the orthogonal
direction. 

\subsection{An ideal filter for circular polarization}
An ideal filter allowing for positive circular polarization is a
device that is totally transparent to a radiation beam whose electric
field component along the $y$-axis has, in a fixed point of space, a
phase lag of 90$^\circ$ with respect to the component along the
$x$-axis, and is totally opaque to a radiation beam which presents a
phase lag of $-90^\circ$ (or a phase anticipation of $90^\circ$)
between the same components. An ideal filter allowing for negative
circular polarization acts in the opposite way. Note that according to
the previous statement, and consistently with our choice of the
right-handed coordinate system $(x,y,z)$, right and left circular
polarization are those defined in Sect.~4.

An ideal filter allowing for positive circular polarization can be
realized by the combination of an ideal quarter-wave plate followed by
an ideal polarizer whose transmission axis is rotated counterclockwise
(looking at the source of radiation) by an angle of $45^\circ$ with
respect to the direction of the fast-axis of the plate.\footnote{
Waveplates are optical elements with two principal axes, one called
the fast-axis, and the other one the slow-axis, characterised by two
different refractive indices. The refractive index of the fast-axis is
smaller than the refractive index of the slow-axis, so that linearly
polarized light with its electric vector parallel to the fast-axis
travels faster than linearly polarized light with its electric vector
parallel to the slow-axis. A quarter-wave plate produces a $\pi/2$
phase retardation between the components of the electric field along
the fast and slow axes.}  An ideal filter allowing for negative
circular polarization can be realized by the combination of an ideal
quarter-wave plate followed by an ideal polarizer whose transmission
axis is rotated clockwise (looking at the source of radiation) by an
angle of $45^\circ$ with respect to the direction of the fast-axis of
the plate.  Note that from the practical point of view it is common to
consider a retarder waveplate free to rotate, followed by a fixed
linear polarizer, the transmission axis of which is aligned to the
reference direction $x$. Obviously, in this configuration, the ideal
filter allowing for positive circular polarization is obtained setting
the fast-axis of the retarder waveplate at an angle of $-45^\circ$
(i.e., $45^\circ$ clockwise) with respect to the reference direction,
looking at the source of radiation, and the ideal filter allowing for
negative circular polarization is obtained setting the fast-axis of
the retarder waveplate at an angle of $45^\circ$ with respect to the
reference direction.

\subsection{The operational definition of Stokes parameters}\label{Sect_Oper}
The operational definition of the Stokes parameters of a radiation
beam can be given in terms of the ideal filters introduced above and
of an ideal detector, capable of producing a signal proportional to
the electromagnetic energy falling on its acceptance area in a given
interval of time and in a given interval of angular frequency
$(\nu, \nu + {\rm d} \nu)$.  We denote by $S$ the signal of
the detector with no filter interposed and by $S_{0}$, $S_{45}$,
$S_{90}$, and $S_{135}$ the signals of the detector with an ideal
filter for linear polarization interposed with the transmission axis
set, respectively, at $0^\circ$, $45^\circ$, $90^\circ$, and
$135^\circ$ with respect to the reference direction, all the angles
being reckoned counterclockwise looking at the source. Finally, we
denote by $S_{+}$ and $S_{-}$ the signals of the detector with an ideal
filter interposed, allowing, respectively, for positive and negative
circular polarization. The Stokes parameters are defined by the
equations

\begin{equation}
\begin{array}{rcl}
I &=& k' S \\
Q &=& k'(S_{0} - S_{90}) \\
U &=& k' (S_{45} - S_{135}) \\ 
V &=& k' (S_{+} - S_{-}) \\
\end{array}
\label{Eq_Oper}
\end{equation}
where $k'$ is a positive constant that turns out to be inessential when considering 
the ratios $P_Q$, $P_U$, and $P_V$ of Eq.~(\ref{Eq_PQ_PU}).

\section{Conclusions}
It can be easily shown that the two definitions of the Stokes
parameters (the one based on the electric field components of
Eq.~(\ref{Eq_Basic}) in Sect.~3, and the operational
one, based on the concept of ideal filters of Eq.~(\ref{Eq_Oper}) in
Sect.~7.3) are completely equivalent in as far as the
fractional Stokes parameters are concerned. The operational definition
is the same as the one proposed by Shurcliff (1962) and both
definitions agree with those of Landi Degl'Innocenti and Landolfi
(2004). We believe that the astronomical community (including
radio-astronomers) would certainly profit in conforming to this
standard. Finally, although each reference direction is as good as any
other in as far as its choice is clearly specified, we propose that,
except for observations of objects belonging to the Solar System, the
astronomical community should conform to the standard of adopting as
reference direction the celestial meridian passing through the
observed object. For solar system objects other than the Sun, we
propose to adopt as reference direction the perpendicular to the
great circle passing through the object and the Sun.


\end{document}